# FrFT based estimation of linear and nonlinear impairments using Vision Transformer

TING JIANG,[1] ZHENG GAO,[1] YIZHAO CHEN,[1] ZIHE HU,[1] AND MING TANG[1,*]

[1] *Wuhan National Laboratory for Optoelectronics (WNLO) and Next Generation Internet Access National Engineering Laboratory (NGIA), School of Optical and Electronic Information, Huazhong University of Science and Technology, Wuhan 430074, China*
*\*tangming@mail.hust.edu.cn*

**Abstract:** To comprehensively assess optical fiber communication system conditions, it is essential to implement joint estimation of the following four critical impairments: nonlinear signal-to-noise ratio ($SNR_{NL}$), optical signal-to-noise ratio (OSNR), chromatic dispersion (CD) and differential group delay (DGD). However, current studies only achieve identifying a limited number of impairments within a narrow range, due to limitations in network capabilities and lack of unified representation of impairments. To address these challenges, we adopt time-frequency signal processing based on fractional Fourier transform (FrFT) to achieve the unified representation of impairments, while employing a Transformer based neural networks (NN) to break through network performance limitations. To verify the effectiveness of the proposed estimation method, the numerical simulation is carried on a 5-channel polarization-division-multiplexed quadrature phase shift keying (PDM-QPSK) long haul optical transmission system with the symbol rate of 50 GBaud per channel, the mean absolute error (MAE) for $SNR_{NL}$, OSNR, CD, and DGD estimation is 0.091 dB, 0.058 dB, 117 ps/nm, and 0.38 ps, and the monitoring window ranges from 0~20 dB, 10~30 dB, 0~51000 ps/nm, and 0~100 ps, respectively. Our proposed method achieves accurate estimation of linear and nonlinear impairments over a broad range, representing a significant advancement in the field of optical performance monitoring (OPM).

## 1. Introduction

To meet the ever-increasing network capacity demands, the optical fiber communication system is becoming more dynamic, flexible, and elastic [1]. However, this transition results in complex impairments in optical signals, necessitating an accurately impairments estimation to maintain quality of transmission (QoT) and margin saving in both current wavelength-division multiplexing (WDM) and future elastic optical networks (EONs) [2-5]. Within the domain of impairments estimation, it is widely accepted the following crucial roles: OSNR, $SNR_{NL}$, CD, and DGD. Firstly, OSNR is the generic benchmark to evaluate the performance of the system [6], therefore monitoring the OSNR remains vital to permit preemptive maintenance and to optimize the link capacity. However, as long-haul transmissions require large launch powers, nonlinear transmission impairments further weaken the signals and introduce large distortions and noise, which cannot be measured by OSNR. The Gaussian-noise (GN) model [7,8] was proposed to calculate $SNR_{NL}$, which represent the accumulated fiber nonlinear interference (NLI) over signal propagation. Therefore, accurately estimating $SNR_{NL}$ and OSNR is especially important. Besides, the first step of the DSP unit at the receiver is CD estimation (CDE), which provides a basis for subsequent CD compensation. However, the CDE is subject to large DGD [9], while DGD itself can cause significant signal degradation at a moment due to statistical characteristics [10]. Therefore, CD and DGD need to be monitored at the same time. To sum up, it is important to achieve joint estimation of OSNR, $SNR_{NL}$, CD, and DGD, leading to cost-effective monitoring at a reasonable algorithm complexity. However, the conventional approach is to tackle each impairment separately, which is highly complex and not very feasible.

Recently, the machine learning (ML) has emerged as a powerful tool in the field of OPM [11], making it possible to achieve joint estimation of multiple impairments. By building maps between received signals and their physical impairments, ML offers advantages such as feature self-extraction, high accuracy, and reliability. Several efforts for joint multi-impairments monitoring by using deep learning (DL) have been proposed over the past years [12-16]. Joint estimation of OSNR, CD and DGD is demonstrated with high precision based on asynchronous delay-tap sampling (ADTS) [12,13]. Despite their high accuracy, the monitoring range of the proposed methods is limited. The CD range is within 500ps/nm and the DGD range is within 0-10ps. Joint estimation of linear and nonlinear SNR based on neural networks is demonstrated with < 0.5 dB MAE for 35-Gbaund DP-16QAM signal over the transmission over 1500-km standard single-mode fiber (SSMF) using features from constellation diagram [14-16]. Although the monitoring of nonlinear impairments has been achieved under long-haul transmission, they can only achieve joint estimation of two types of impairments, while the accuracy is not high enough. In summary, current studies are insufficient to satisfy the requirements as they only identify a limited number of impairments within a narrow range. Up until now, no method has been able to achieve high-precision, wide-range joint estimation of the four aforementioned impairments, due to two main reasons:

1. Limited network ability. There is mutual restraint between the accuracy of the algorithm and the complexity of the network. A more powerful network structure is urgently needed to achieve high-precision, wide-ranging, and low-complexity integrated monitoring of multiple impairments.

2. Input features lack a unified representation of multiple impairments. Currently, in the research of OPM based on NNs, the input features of the network are complex and inconsistent, which limits the types of impairments that can be monitored by a single feature, making it difficult to achieve integrated monitoring of multiple impairments.

Regarding the problem one, Transformer [17], an attention-based encoder-decoder architecture, has revolutionized the field of natural language processing, computer vision, and audio processing. Current works show that Transformer-based pre-trained models (PTMs) [18] can achieve state-of-the-art performances on various tasks. For example, the Generative Pre-trained Transformer (GPT) families [19,20] has pushed the boundaries of natural language generation and understanding, while Vision Transformer (ViT) [21] outperforms traditional CNNs on mainstream classification benchmarks. Currently, Transformer has not been well explored and utilized in OPM. Inspired by the recent progress of Transformer, applying Transformer to the field of OPM is promising for achieving higher precision and wider range of joint estimation of multiple impairments.

Regarding the problem two, we believe that the joint time-frequency signal processing method based on FrFT can achieve unified representation of multiple impairments. Previous researches only processed signals in either a single time domain or frequency domain perspective, which cannot represent the relationship between signal frequency components and time variation. To overcome the limitations of traditional Fourier analysis, a time-frequency domain signal processing method can be used to represent signals using joint functions of time and frequency. We previously demonstrated a joint CD, DGD and PDL estimation algorithm using FrFT based time-frequency distribution reconstructed images and multi-task CNN [22]. However, this work is focused on linear impairments monitoring in single carrier coherent scenarios, where most of the fiber nonlinearity comes from the self-phase modulation (SPM) effect. In WDM systems, the inter-channel cross-phase-modulation (XPM) constitutes the predominant contribution to nonlinear noises. Therefore, it is necessary to further monitor the performance of nonlinearity impairments in WDM systems.

In this paper, we combine the advantages of the time-frequency signal processing based on FrFT and the Transformer based multi-task ViT to achieve joint estimation of OSNR, $SNR_{NL}$, CD, and DGD. To our knowledge, this is the first application of Transformer in the OPM field. We use FrFT to comprehensively observe and process signals by reconstructing the time-frequency distribution images, which are then used to train ViT and extract features that represent impairments, ultimately enabling joint estimation of OSNR, SNR$_{NL}$, CD, and DGD. To validate the effectiveness of our proposed method, we conduct numerical simulations on a 5-channel PDM-QPSK long-haul optical fiber transmission system with a symbol rate of 50GBaud per channel, and transmission distances ranging from 100 to 3000 km. The impairment ranges consider were: 10~30dB for OSNR, 0~20dB for SNR$_{NL}$, 1700~51000ps/nm for CD, and 0~100ps for DGD. Our proposed method achieves a MAE of 0.058 dB, 0.091 dB, 117 ps/nm, and 0.38 ps, respectively, outperforming existing OPM techniques. Furthermore, we conduct a comparison of our proposed ViT-based method with other NN-based methods and verify its superior algorithmic precision. We also evaluate the performance of our approach under different orders of FrFT.

## 2. Operation Principle

In this section, we first introduce the relationship among SNR$_{NL}$, OSNR, and general signal-to-noise ratio (GSNR). Then, we provide a detailed explanation on cascade PMD model in real-world scenarios, including mathematical derivation and modeling process. Finally, we describe the network architecture and mathematical principles of ViT in detail, and demonstrate the multi-impairments joint estimation method based on ViT.

### 2.1 SNR$_{NL}$, OSNR and GSNR

To accurately evaluate the quality of a long-haul WDM transmission system, it is essential to consider both the effect of ASE noise and nonlinearity. Therefore, the GSNR is proposed to quantify their combined impact and represent the overall quality of the channel [23,24], which can be expressed as:

$$GSNR[dB] = 10\log_{10}(\frac{P_{sig}}{P_{NL} + P_{ASE}}) \tag{1}$$

where P$_{sig}$, P$_{ASE}$, and P$_{NL}$ represent the electrical power of signal, ASE noise, and NLI noise, respectively. Regarding nonlinear impairments, SNR$_{NL}$ represents the ratio of received signal power to the NLI noise power, providing valuable insights into optimizing network design and enhancing system performance [25]. The SNR$_{NL}$ is defined as:

$$SNR_{NL}[dB] = 10\log_{10}(\frac{P_{sig}}{P_{NL}}) \tag{2}$$

OSNR is defined as the ratio of the signal power and the amplified spontaneous emission (ASE) noise within the signal bandwidth, which is generated by the erbium-doped fiber amplifier (EDFA) during the amplification, which can be expressed as:

$$OSNR[dB] = 10\log_{10}(\frac{P_{sig}}{P_{ASE}}) \tag{3}$$

where P$_{sig}$ is the optical signal power and P$_{ASE}$ is the ASE noise power within 0.1-nm measurement bandwidth. The fundamental relationship can be established between SNR$_{ASE}$ and the OSNR [26] as follows:

$$SNR_{ASE} = \frac{B_{ref}}{R_s}ONSR \tag{4}$$

Where Rs is the symbol rate, and B$_{ref}$ denotes the reference bandwidth which is usually treated as 12.5 GHz. Hence, the following Eq. (5) can be used to express the relationship among SNR$_{NL}$, SNR$_{ASE}$, and GSNR:

$$GSNR^{-1} = SNR_{NL}^{-1} + SNR_{ASE}^{-1} \tag{5}$$

### 2.2 Cascaded PMD model in fiber

In the case of short distance, the birefringent fiber can behave as a waveplate and the PMD is no longer statistical. However, under the situation of long-distance transmission, PMD is time-varying and statistically. Therefore, we adopt a cascaded model consisting of multiple birefringent fiber segments to accurately simulate real-world cases of PMD in the fiber [27]. We simulate the random birefringence in the fibers by a cascade of waveplates with randomly oriented axes, where the DGD generated by the $i_{th}$ small segment of fiber is $\Delta\tau_i(l_i)$, and the slow axis orientation is $\alpha_i$, which is shown in Fig. 1. and Eq. (6) as follows:

$$\Delta\tau_i(l_i) = \sqrt{\frac{3\pi l_i}{8}} D_{PMD} \quad (6)$$

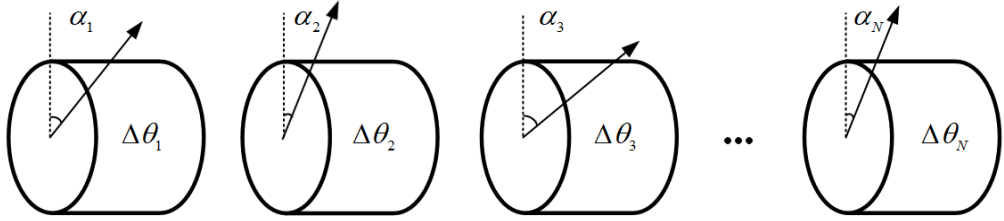

Fig. 1. Cascaded PMD model of randomly birefringent fiber.

where $D_{PMD}$ represent PMD parameter and $l_i$ is the length of $i_{th}$ segment of the fiber. To generate PMD in the range of 0~100 ps, we employed 20 DGD segments, each containing randomly rotating Jones matrices. The final Jones transmission matrix of the entire optical fiber can be expressed as:

$$U_{PMD}(\omega) = \prod_{i=i}^{N} \begin{bmatrix} \cos\alpha_i & -e^{-j\gamma_i}\sin\alpha_i \\ e^{j\gamma_i}\sin\alpha_i & \cos\alpha_i \end{bmatrix} \begin{bmatrix} \exp(\frac{j\omega\Delta\tau_i}{2}) & 0 \\ 0 & \exp(-\frac{j\omega\Delta\tau_i}{2}) \end{bmatrix} \begin{bmatrix} \cos\alpha_i & -e^{-j\gamma_i}\sin\alpha_i \\ e^{j\gamma_i}\sin\alpha_i & \cos\alpha_i \end{bmatrix}^{-1} \quad (7)$$

Considering that the transmission matrix can be expressed in the form of Caley-Klein parameterization:

$$U_{PMD} = \begin{bmatrix} u_1 & u_2 \\ -u_2^* & u_1^* \end{bmatrix} \quad (8)$$

Hence, the DGD of the entire fiber, denoted by $\Delta\tau$, is calculated by the following formula:

$$\Delta\tau = 2\sqrt{|u_{1w}|^2 + |u_{2w}|^2} \quad (9)$$

where $u_{1w}$ and $u_{2w}$ represent the derivatives of $u_1$ and $u_2$ with respect to the angular frequency. The distribution of first-order PMD in optical fibers follow a Maxwellian distribution:

$$P(x = \Delta\tau) = \frac{32x^2}{\pi^2 <\Delta\tau>^2} \exp\left[-\left(\frac{4x^2}{\pi <\Delta\tau>^2}\right)\right] \quad (10)$$

where $<\Delta\tau>$ represents the average DGD. The distribution diagram of DGD is shown below:

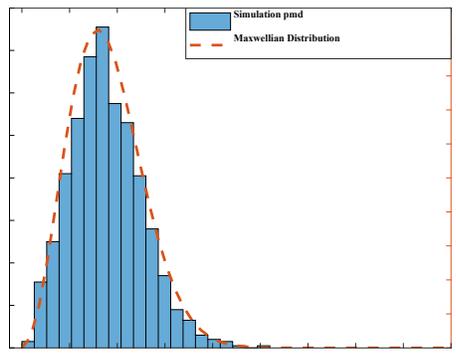

Fig. 2. The distribution diagram of DGD which follows Maxwellian distribution.

*2.3 Vision Transformer*

The Transformer model architectures have gained much attention due to their remarkable performance across diverse domains, including natural language processing (NLP), computer vision (CV), and reinforcement learning. Compared with conventional deep learning models such as CNN [28] and LSTM [29], Transformer completely dispenses with recurrence and convolutions by employing multi-head self-attention (MHSA) mechanisms and point-wise feed-forward networks (FFNs). Inspired by the undeniable success of Transformers in NLP tasks, Vision Transformer (ViT) is proposed by directly applying a standard Transformer architecture to images, which surpasses most conventional CNN-based methods across multiple image recognition benchmarks.

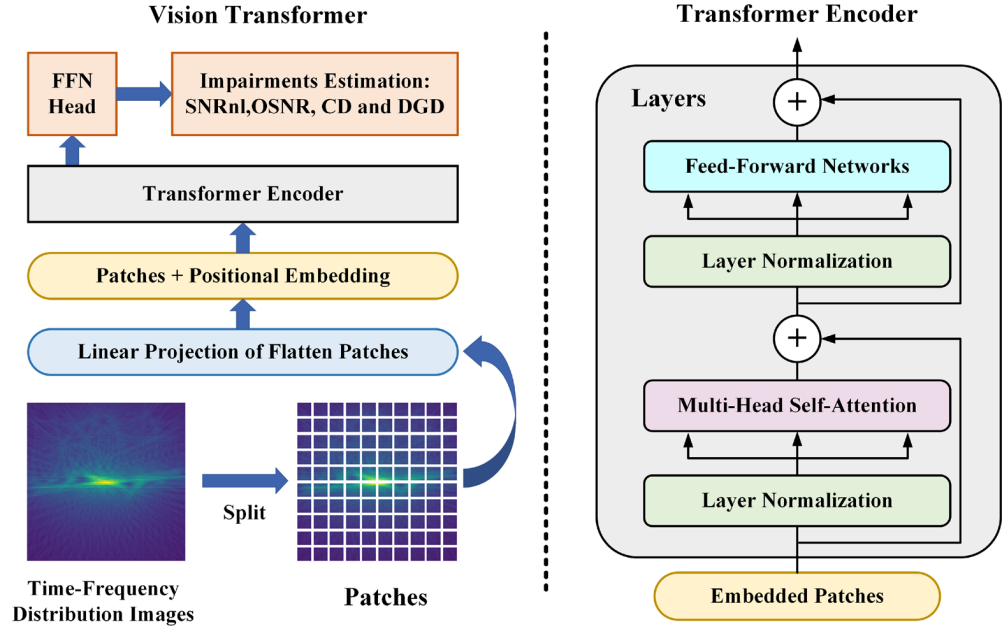

Fig. 3. Model architecture of Vision Transformer.

An overview of the network architecture is shown in Fig. 3. The input consists of a time-frequency distribution image obtained by combining FrFT at different orders with the inverse Radon transform, and the specific process will be elaborated in the next section. Its dimension size is 100 × 100 × 2, where 100 represents the length and width of the image, and 2 represents x and y polarization states. Initially, the input image is split into a series of non-overlapped patches that are subsequently projected into linear patch embeddings. Transformer doesn't utilize recurrences or convolutions, resulting in losing sequential order information. Therefore, a positional encoding is added to each patch to preserve the spatial information, then the joint embeddings are fed into the Transformer encoder. The Transformer encoder has two sub-layers. The first is a MHSA mechanism, and the second is a simple, position-wise fully connected FFN. To build a deep model, a residual connection [30] is employed around each of the two sub-layers, followed by layer normalization [31]. Therefore, the Transformer encoder block is written as:

$$X_1 = MultiHead(LayNorm(X)) + X$$
$$X = FFN(LayNorm(X_1)) + X_1 \quad (11)$$

The core component of Transformer is the self-attention mechanism with Query-Key-Value (QKV) model. Specifically, given the matrix representations of queries $Q \in R^{N \times D_k}$, keys $K \in R^{M \times D_k}$, and values $V \in R^{M \times D_v}$, the scaled dot-product attention used by Transformer is:

$$Attention(Q,K,V) = Soft\max(\frac{QK^T}{\sqrt{d_k}})V \quad (12)$$

Where the scaling factor $d_k$ and a SoftMax operation are applied to normalize the attention weights into a probability distribution, which helps to avoid gradient vanishing problem during training. Instead of performing a single attention function, Transformers adopt MHSA mechanism to linearly project the input into multiple subspaces, which are processed in parallel by multiple independent attention heads. All the outputs are concatenated and projected to the output representation, as depicted in Fig. 4.

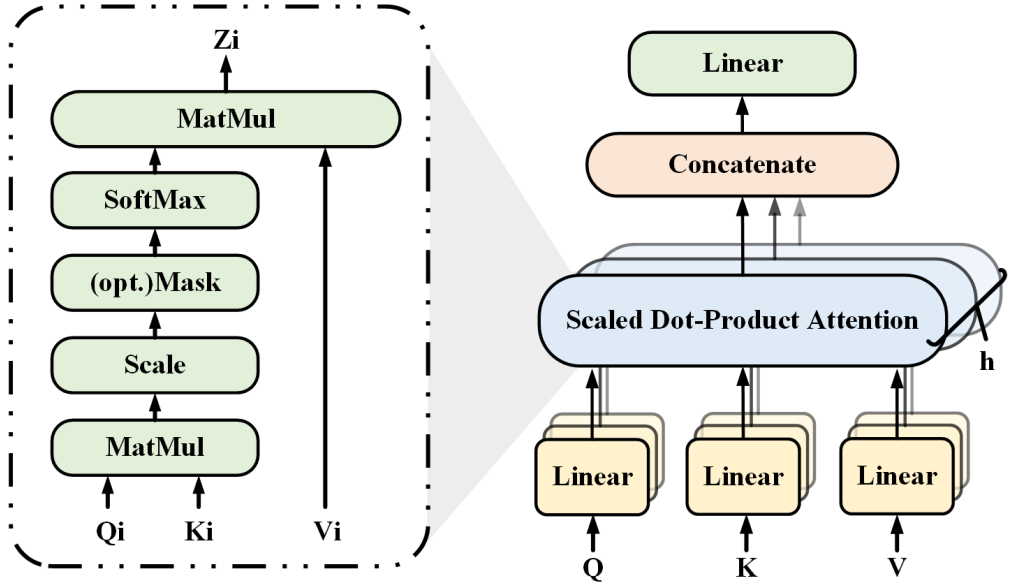

Fig. 4. The structure of the MHSA layer. Left: Scaled Dot-Product Attention. Right: Multi-Head Attention Mechanism.

The process of MHSA can be formulated as:

$$MultiHead(Q,K,V) = ConCat(head_i,...,head_h)W^O \quad (13)$$
$$head_i = Attention(QW^{Q_i}, KW^{K_i}, VW^{V_i}), i = 1,2,...,h \quad (14)$$

where the projections are parameter matrices $W_i^{Q_i} \in R^{D_{model} \times D_k}$, $W_i^{K_i} \in R^{D_{model} \times D_k}$, $W_i^{V_i} \in R^{D_{model} \times D_v}$ and $W^O \in R^{hD_v \times D_{model}}$, while h is the number of heads. In this work we employ h = 4 parallel attention heads. For each of these we use $D_k = D_v = D_{model} / h = 64$. The output of MHSA is then fed into two successive FFNs with a rectified linear unit (GeLU) non-linearity activation as:

$$FFN(x) = GELU(W_1 x + b_1)W_2 + b_2 \qquad (15)$$

After passing through the Transformer Encoder layer, the compressed features are flattened into a 1D vector, which then undergo FFN to conduct four regression tasks: OSNR estimation, $SNR_{NL}$ estimation, CDE estimation, and DGD estimation. The mean absolute error (MAE) is used as the corresponding loss function for the impairment estimation output, as it outperforms root mean square error (RMSE) in evaluating model accuracy [32]:

$$L_{MAE} = \frac{1}{N}\sum_{i=1}^{N}|a_i - \hat{a}_i| \qquad (16)$$

Thus, the final loss function used to evaluate the convergence of MT-ViT during training can be expressed as:

$$Loss_{all} = \lambda_1 L_{CD} + \lambda_2 L_{DGD} + \lambda_3 L_{OSNR} + \lambda_4 L_{SNR_{nl}} \qquad (17)$$

where the parameter $\lambda_i$ is utilized to adjust the relative importance of various tasks in the MT- ViT model, thereby enhancing its overall performance. The model is trained using the adaptive gradient momentum (Adam) optimization algorithm, with the goal of minimizing output errors [33].

## 3. Simulation Setup

To evaluate the effectiveness of our proposed method, we conducted a comprehensive numerical simulation using commercial software Virtual Photonics Inc. (VPI) Transmission Maker, MATLAB, and Keras library over a wide range of system configurations. Fig. 5 and Fig. 6 illustrate the schematic diagram of linear and nonlinear impairments estimation method based on FrFT and ViT network for a 5-channel long-haul PDM-QPSK coherent WDM optical transmission system, while Table 1 presents the primary fiber channel parameters utilized in this work.

At the transmitter (Tx) side, two independent random sequences generated in Matlab are used to modulate two IQ Modulators at 100GSam/s to form a 50Gbaud PDM-QPSK signal superimposed with 0.1 order TS, while root-raised-cosine (RRC) pulse shaping with a roll-off factor of 0.02 was applied to perform shaping. The length of each symbol sequence and TS is $2^{14}$ and 100, respectively. The launched power per channel is ranged from 0 to 6 dBm. The channel spacing is 75 GHz and the center channel is under the test.

The modulated optical signal is amplified by an EDFA and launched into a recirculating loop, each span consists of a 100 km long SSMF with an attenuation coefficient of 0.2 dB/km and an EDFA to compensate for the transmission loss of each span. The simulation considers the SPM as well as the XPM effects caused by the fiber nonlinearity with a nonlinear coefficient of 2.6 $W^{-1} \cdot km^{-1}$. The CD parameter of SSMF is 17 ps/ (nm × km) and one loop accumulates 1700 ps/nm CD. The noise figure of the EDFA in loop is set to 0 dB and the OSNR is controlled by adding ASE noise at the end of the fiber link.

After exiting the loop, PMD is emulated by cascading 20 randomly rotating Jones matrices, with a range of 0 to 100 ps. The DGD of the PMD matrices is considered for determining the PMD value, which follows the Maxwellian probability distribution. At the receiver end, an optical band-pass filter (OBPF) with a bandwidth of 75 GHz is used to filter out the optical signal before it is detected by a coherent receiver for further DSP.

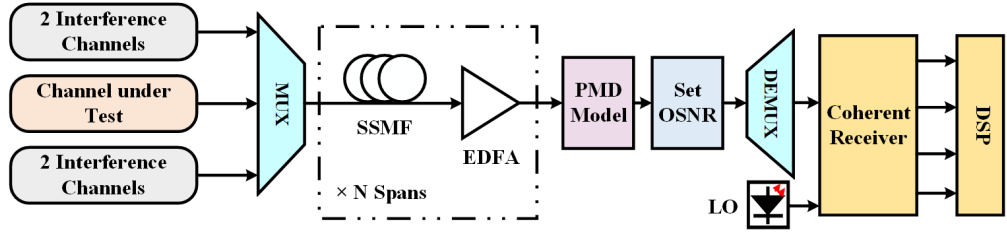

Fig. 5. Simulation setup of 5-channel PDM-QPSK WDM optical transmission system.

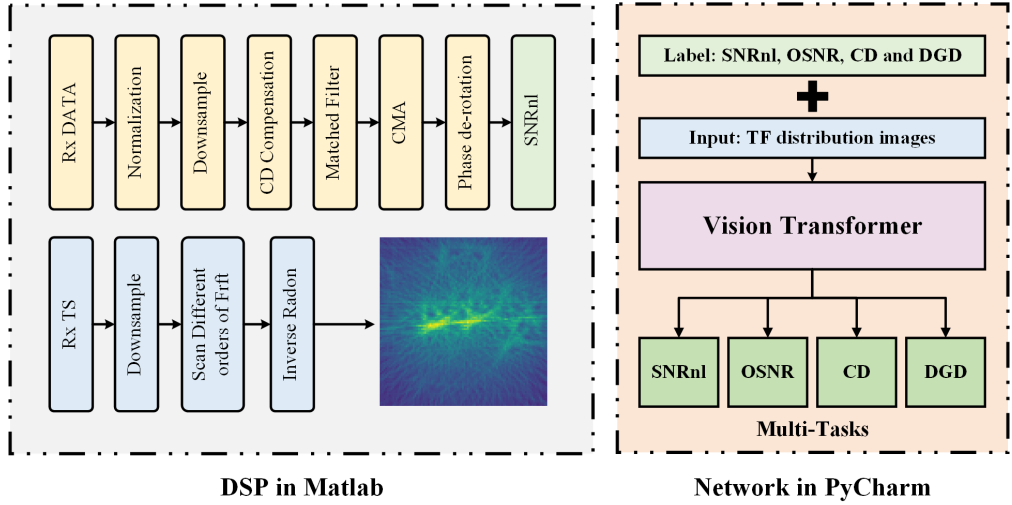

Fig. 6. Schematic diagram of ViT based multi-impairments estimation method for WDM system.
Left: DSP process in MATLAB.  Right: Network training process in PyCharm.

The received signals are processed by normalization, down-sampling, chromatic dispersion compensation, matched filtering, constant modulus algorithm (CMA) adaptive equalization, and phase de-rotation after coherent detection. The transmitted signal and the received signal are combined to calculate GSNR. Based on Eq. (4), OSNR can be converted to $SNR_{ASE}$. Finally, the target $SNR_{NL}$ of the proposed method can be obtained by Eq. (5). The $SNR_{NL}$, OSNR, CD and DGD are treated as the benchmark in the following part to evaluate the accuracy of impairments estimation.

The received TSs don't need to go through the complete DSP process, which are resampled and performed FrFT scanning at different orders ranging from -1 to 1, with an interval of 0.05, the definition of FrFT is as follows:

$$X_\alpha(u) = \{F^\alpha[x(t)]\}(u) = \int_{-\infty}^{+\infty} f(t) K_\alpha(t,u) dt$$

$$= \begin{cases} \int_{-\infty}^{+\infty} \sqrt{\frac{1-j\cot\alpha}{2\pi}} \exp(j\frac{t^2+u^2}{2}\cot\alpha - \frac{jtu}{\sin\alpha}) x(t) dt & \alpha \neq n\pi \\ x(t) & \alpha = 2n\pi \\ x(-t) & \alpha = (2n\pm 1)\pi \end{cases} \quad (18)$$

where $\alpha$ is the rotation angle, the order of FrFT is defined as $2\alpha/\pi$. From the equation we can see that FrFT has the rotation property, thus different orders of FrFT scanning can obtain time-frequency distribution images in polar coordinates. Then combining with the Radon inverse transformation, the back-projection is carried out by linear interpolation. Finally, the time-frequency images in standard rectangular coordinate system are reconstructed with a size of 100 × 100×2, then imported into MT-ViT to conduct impairments estimation.

Table 1. Summary of link configurations of WDM optical fiber communication system

| Parameters | Value | Parameters | Value |
| --- | --- | --- | --- |
| Modulation format | PDM-QPSK | Fiber type | SSMF |
| Symbol length | $2^{14}$ | Attenuation | 0.2dB/km |
| TS length | 100 | Nonlinear coefficient | 2.6 W−1 · km−1 |
| Roll-off factor | 0.02 | Dispersion parameter | 17 ps/ (nm × km) |
| Symbol rate | 50GBaud/ channel | Span length | 100km |
| Sampling rate | 2 Symbols/ Sample | Span | 1~30 |
| Channel spacing | 75GHz | DGD | 1~100 ps |
| Lauched power | 0~6 dBm/ channel | OSNR | 10~30 dB |

We collect 12,000 sets of data based on the aforementioned system, which are divided into three parts: 80% for the training set (9600 samples), 10% for the validation set (1200 samples), and 10% for the test set (1200 samples), respectively. To establish our MT-ViT method, we utilize the PyCharm platform and the Keras library with Tensorflow as the backend on a personal computer equipped with a graphics processing unit (GPU) accelerator during the training and testing processes. For hyperparameter tuning, Bayesian optimization [34] is employed to perform efficient exploration of high-dimensional search spaces by updating the weight and bias values, which saves considerable time compared with random searching or grid searching. During the training phase, we monitored the model's performance on both the training and validation datasets, and adjusted its parameters to avoid overfitting. Once the model achieves satisfied performance and convergence, it is evaluated on the test dataset to assess its generalization ability.

## 4. Simulation results

In this chapter, we first discuss the performance of the proposed linear and nonlinear impairments estimation method under different ASE noise and fiber nonlinearity variations. Then, we compare the performance of the proposed method under different network frameworks. Finally, we discuss effectiveness under different time-frequency distribution images established at different FrFT orders.

### 4.1 Performance of the proposed method

After the simulation, we obtained time-frequency distribution images and corresponding impairments labels using Matlab, which were subsequently fed into ViT for multi-tasking supervised learning. After optimizing the hyperparameters, we conducted five rounds of training with the optimal settings to mitigate the impact of model initialization variability and presented the best-performing model.

Table 2. Performance of proposed method with launch power of 2 dBm/channel

| | Estimation Range | MAE |
| --- | --- | --- |
| $SNR_{NL}$ estimation | 0~20 dB | 0.085 dB |
| OSNR estimation | 10~30 dB | 0.060 dB |
| CD estimation | 0~50400 ps/nm | 120.3 ps/nm |
| DGD estimation | 0~100 ps | 0.33 ps |

Our simulation results are presented from the perspectives of long-haul transmission distances, different launch powers, and varying ASE noises. Fig. 7. (a)-(d) show the simulation results for joint estimation of $SNR_{NL}$, OSNR, CD, and DGD under transmission conditions with five channels at 2 dBm/channel. The figure illustrates that the blue dashed line corresponds to the estimated value, while the black solid line represents the fitted curve. The red solid line shows the segmental MAE for different interval values. The specific numerical performance value is shown in Table 2. The estimation range for $SNR_{NL}$, OSNR, CD and DGD is 0~20dB, 10~30dB, 0~50400ps/nm, and 0~100ps, respectively. Their corresponding MAE for estimation is 0.085dB, 0.060dB, 120.3ps/nm, and 0.33ps, respectively. Based on the above figure, it can be confirmed that the proposed estimation method is both feasible and sufficiently accurate.

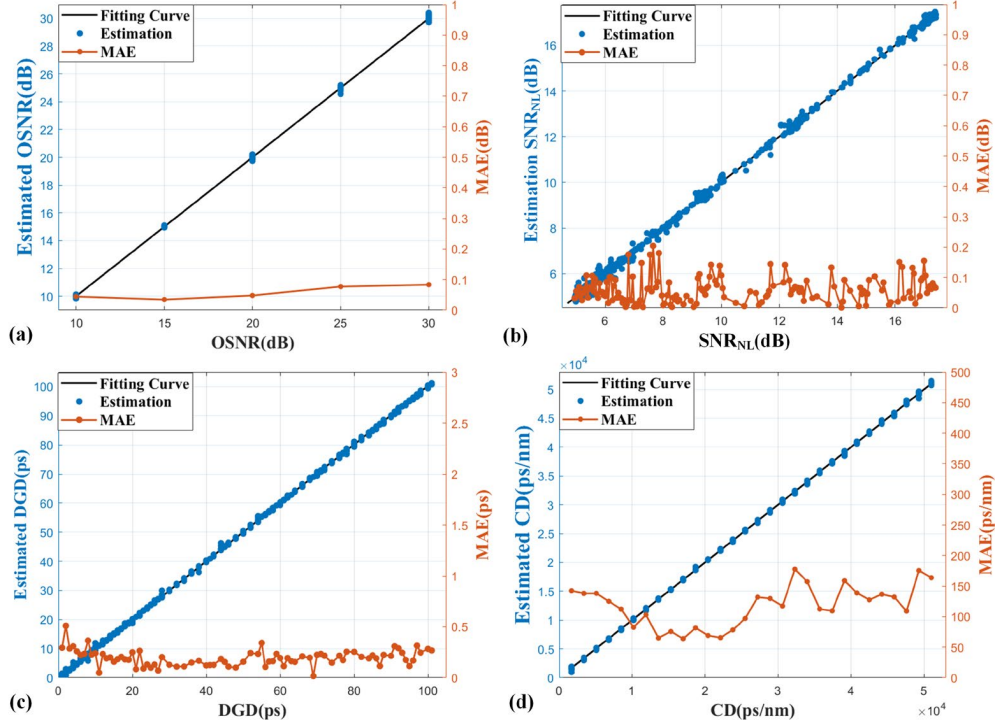

Fig. 7. Performance of the proposed method. (a) OSNR estimation. (b) $SNR_{NL}$ estimation. (c) DGD estimation. (d) CD estimation.

The presence of fiber nonlinearity and ASE noise can affect the joint estimation of linear and nonlinear impairments. To verify that the proposed method can achieve accurate estimation of multiple impairments under complex ASE noise and nonlinear interference, we simultaneously scanned the OSNR from 10 to 30 dB and input signal power from 0 to 6 dBm. The numerical performance values are summarized in Table 3. The total estimation MAE for $SNR_{NL}$, OSNR, CD and DGD is 0.091dB, 0.058dB, 117.3ps/nm, and 0.38ps, respectively. Based on these results, we confirm that our method is indeed robust to ASE noise and fiber nonlinearity. Based on the above system, we can effectively verify that the proposed method can achieve high-precision and large-scale impairments monitoring, while efficiently identifying linear and nonlinear impairments.

**Table 3. Performance under varying fiber nonlinearity and ASE noise.**

|  | Pin=0dBm | Pin=2dBm | Pin=4dBm | Pin=6dBm | Total |
|---|---|---|---|---|---|
| $SNR_{NL}$ estimation | 0.089 dB | 0.085 dB | 0.112 dB | 0.081 dB | 0.091 dB |
| OSNR estimation | 0.057 dB | 0.060 dB | 0.060 dB | 0.056 dB | 0.058 dB |

| CD estimation | 114.3 ps/nm | 120.3 ps/nm | 117.7 ps/nm | 116.2 ps/nm | 117.3 ps/nm |
| DGD estimation | 0.39 ps | 0.33 ps | 0.47 ps | 0.34 ps | 0.38 ps |

*4.2 Performance comparison of neural networks with different architectures*

The impact of network architecture on the accuracy of estimation is significant, therefore it is necessary to explore the estimation performance under different NN architectures. On the one hand, CNN is effective in processing two-dimensional image data, enabling efficient extraction of spatial features through convolution while effectively downsampling the data with pooling. On the other hand, DNN excels at processing one-dimensional information, allowing for low-complexity fitting regression tasks to be performed with high accuracy. To further demonstrate the effectiveness of MT-ViT, we conducted a detailed comparison between MT-ViT, MT-CNN, and MT-DNN on the same dataset.

The architecture of MT-CNN is shown in the Fig. 8. (a). Our CNN network architecture includes one batch normalization (BN) layer, four convolutional layers, two pooling layers, one flatten layer, and two dense layers. Firstly, the input time-frequency distribution images go through a BN layer to normalize the activations of the input data, thereby avoiding vanishing gradients. The input then passes through two convolutional layers that apply a set of filters to slide over the input tensor and extract deep and complex features with certain local patterns. After the second convolutional layer, a pooling layer is introduced to reduce the spatial dimensions while retaining important information, which helps to increase the network's robustness and generalization ability. The output subsequently goes through two more sets of alternating convolutional and pooling layers. Finally, it is flattened into a one-dimensional vector and passed through two dense layers for the final output mapping. The structure of MT-DNN is illustrated in Fig. 8. (b). Firstly, the time-frequency distribution images are stretched into one-dimensional vectors. Then, the input passes through a BN layer and six dense layers sequentially to extract the corresponding features of the impairments. Finally, multi-task training is achieved by parallel adding four dense layers with a dimension of 1.

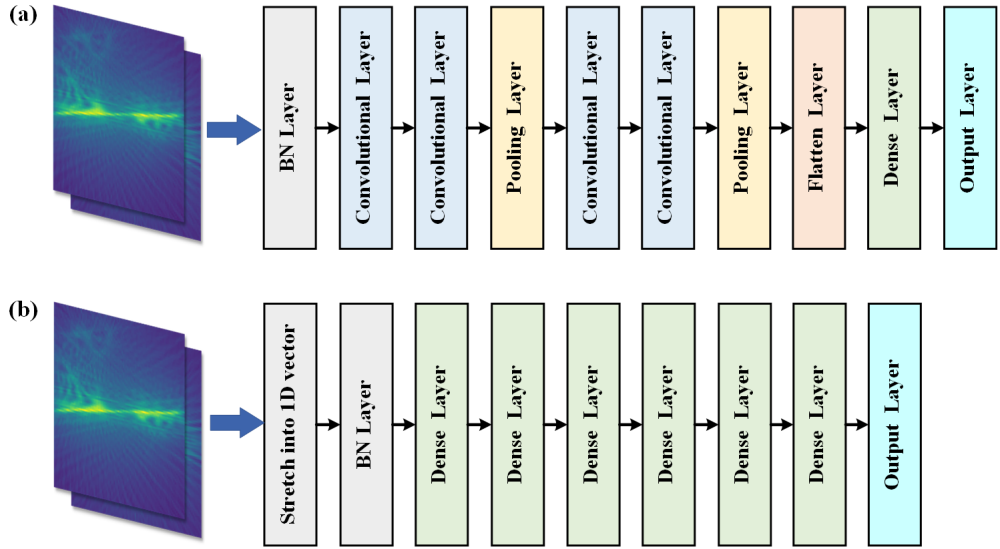

Fig. 8. The architecture of comparison NN. (a) MT-CNN. There are 4 convolutional layers in total. The first and third use 32 and 64 filters sized 5x5, while the second and fourth have 32 and 64 filters sized 3x3, respectively. All pooling layers are MaxPooling and there are 256 neurons in the dense layer. (b) MT-DNN. The dense layers consist of seven different neuron sizes: 2000, 800, 400, 100, 50, 20, and 10, respectively. The other network parameters are completely consistent with the ViT settings.

We train and test the networks on the same dataset and summarize their performance in Fig. 9 and Table 4. Fig. 9 indicates that CNN and DNN have slower convergence speed compared to ViT, and their loss functions do not decrease with the increase of epoch in the later training stages (epoch>4000), suggesting they have reached a local minimum point. In contrast, ViT is capable of further optimizing the parameters and reducing the loss function, implying that the ViT better captures global contextual information by incorporating self-attention mechanisms across all patches of an image, therefore its network has higher capacity compared to CNN and DNN. The results from Table 4 demonstrate that ViT significantly outperforms CNN and DNN in terms of accuracy, indicating that the ViT network can achieve high-accuracy, large-scale, and multi-impairments monitoring based on time-frequency distribution images.

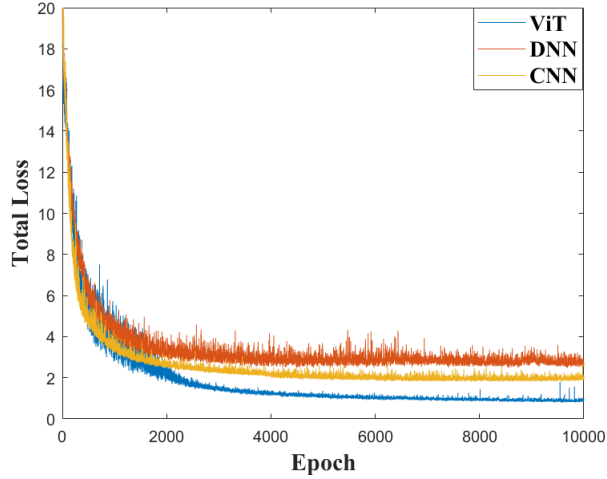

Fig. 9. The curves of the loss function for different networks.

**Table 4. Performance of different models on the same dataset**

|  | MAE of ViT | MAE of CNN | MAE of DNN |
| --- | --- | --- | --- |
| $SNR_{NL}$ estimation | 0.091 dB | 0.21 dB | 0.37 dB |
| OSNR estimation | 0.058 dB | 0.19 dB | 0.27 dB |
| CD estimation | 117.3 ps/nm | 314.5 ps/nm | 415.9 ps/nm |
| DGD estimation | 0.38 ps | 0.84 ps | 1.04 ps |

*4.3 Performance Under Different Orders of TS*

The previous section discussed the impact of different network structures on the performance of the OPM method. In this section, we will discuss the performance with different input features. Since different orders of FrFT of TS occupy different frequency range, the order of the TS is very important as it affects the distribution characteristics of the time-frequency distribution images. High order occupies a large range of frequency domain, which is shown in the Fig. 10. To explore the performance of the proposed method under different inputs, we scanned the order of TS from 0.1 to 0.9 while keeping other simulation settings unchanged and trained the VIT network under the same configuration.

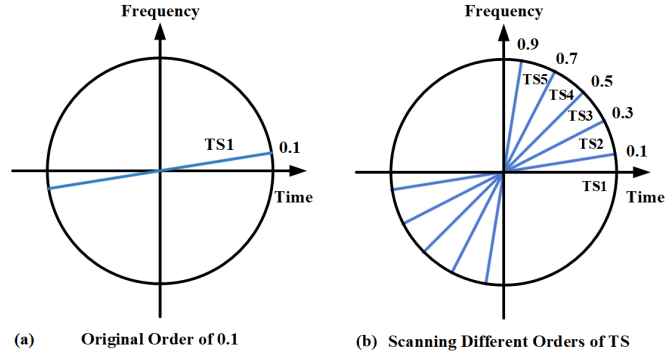

Fig. 10. The architecture of comparison NN. (a) MT-CNN. (b) MT-DNN. Diagram of time-frequency distribution for TS

The specific performance values are summarized in the Table 5. Based on the results, it can be concluded that there is no significant difference in algorithm performance under different TS orders. This is due to our simulation scheme approximating the mainstream trunk parameter configuration in the real scenario, with channel symbol rate of 50GBaud and channel spacing of 100GHz. However, in future high-bandwidth WDM systems, with increasing system symbol rate, high-order PMD impairments will become more prominent, thus making the impact of the order on performance more significant.

Table 5. Performance under different orders of FrFT

| Order of FrFT | $SNR_{NL}$ estimation | OSNR estimation | CD estimation | DGD estimation |
| --- | --- | --- | --- | --- |
| 0.1 | 0.091dB | 0.058dB | 117.3ps/nm | 0.38ps |
| 0.3 | 0.101dB | 0.076dB | 166.6ps/nm | 0.82ps |
| 0.5 | 0.097dB | 0.071dB | 219.1ps/nm | 0.78ps |
| 0.7 | 0.101dB | 0.052dB | 141.1ps/nm | 0.48ps |
| 0.9 | 0.099dB | 0.06dB | 138.1ps/nm | 0.50ps |

## 5. Conclusion

In this paper, we propose a joint $SNR_{NL}$, OSNR, CD and DGD estimation method base on FrFT and Vision Transformer. In order to address the problem of lack of unified representation of impairments in the OPM field, we adopt time-frequency signal processing based on FrFT to comprehensively observe and process signals from a time-frequency two-dimensional perspective. To solve the problem of overfitting in complex multi-impairments situations for the neural network, we introduce Transformer to achieve multiple impairments estimation. A 5-channel long-haul PDM-QPSK coherent WDM system is established to validate the feasibility of the method. The simulation results demonstrate that by analyzing impairments from the perspective of time-frequency distribution images and incorporating Transformer network, high-accuracy and broad-range estimation of linear and nonlinear impairments can be achieved. This approach is not only essential in the current dominant WDM systems but also holds considerable potential for future EON and intelligent network operations.

**Funding.** National Natural Science Foundation of China (61931010, 62225110).

**Disclosures.** The authors declare no conflicts of interest.

**Data availability.** Data underlying the results presented in this paper are not publicly available at this time but may be obtained from the authors upon reasonable request.